\documentstyle[aps,twocolumn]{revtex}
\begin{document}
\input{psfig}
\draft
\title{Positron-neutrino correlation in 
the $0^+ \rightarrow 0^+$ decay of $^{32}$Ar}
\author{E.~G.~Adelberger$^1$,  C. Ortiz$^2$, A.~Garc\'{\i}a$^2$,
H.E. Swanson$^1$, M. Beck$^1$, O. Tengblad$^3$, M.J.G. Borge$^3$, 
I. Martel$^4$, \\ 
H. Bichsel$^1$ and the ISOLDE collaboration$^4$}
\address{$^1$Department of Physics, University of Washington, 
Seattle, Washington 98195-1560}
\address{$^2$Department of Physics, University of Notre Dame, 
Notre Dame, Indiana 46556}
\address{$^3$Instituto de Estructura de la Materia, 
CSIC, E-28006 Madrid, Spain}
\address{$^4$EP Division, CERN, Geneva, Switzerland CH-1211}
\date{\today}
\maketitle
\begin{abstract}
The positron-neutrino correlation in the $0^+ \rightarrow 0^+$ 
$\beta$ decay of $^{32}$Ar was measured at ISOLDE by analyzing the effect 
of lepton recoil on the shape of the narrow proton group following 
the superallowed decay. 
Our result is consistent with the Standard Model prediction.
For vanishing Fierz interference we find 
$a=0.9989 \pm 0.0052 \pm 0.0036$, which yields
improved constraints on scalar weak interactions.
\end{abstract}
\begin{narrowtext}
The $e^+$-$\nu$ correlation in $0^+ \rightarrow 0^+$ $\beta$ decay
provides a robust signature of possible rare processes where particles other
than the usual $W^{\pm}$ boson are exchanged. In the Standard Model
and its 
left-right symmetric generalizations, $0^+ \rightarrow 0^+$ decays
produce the $e^+$ and $\nu$ with opposite chiralities.
Angular momentum conservation then prevents relativistic leptons
from being emitted in {\em opposite} directions.
On the other hand, a scalar interaction, 
which could arise from exchange of
a leptoquark or a Higgs boson\cite{he:95}, will produce the
$e^+$ and $\nu$ with identical chiralities so that, in the 
relativistic limit, the leptons cannot be emitted in {\em the same} direction.
In minimal extensions of the Standard Model, Higgs couplings are 
too small to affect significantly the $e$-$\nu$ correlation.
However, supersymmetric theories with more than one charged Higgs 
doublet can accommodate sizable scalar couplings \cite{he:95}
that are not ruled out by existing data \cite{bo:84}.

This paper reports new constraints on scalar weak interactions based
on a precise measurement of the $e$-$\nu$ correlation
in the $0^+ \rightarrow 0^+$ $\beta^+$ decay of $^{32}$Ar, the only pure
Fermi transition whose $e$-$\nu$ correlation has been determined with good
precision.
The simple spin structure of this decay permits tests for scalar
interactions without complications from axial or tensor currents or from
recoil-order effects.
We consider a $0^+ \rightarrow 0^+$ $\beta^+$-decay Hamiltonian\cite{ja:57} 
\begin{eqnarray}
H = (\bar{\psi_n} \gamma_\mu \psi_p)
         (C_V  \bar{\psi_{\nu}} \gamma_\mu \psi_e + 
          C_V' \bar{\psi_{\nu}} \gamma_\mu \gamma_5 \psi_e)+  \nonumber\\
    (\bar{\psi_n} \psi_p)
         (C_S  \bar{\psi_{\nu}} \psi_e + 
          C_S' \bar{\psi_{\nu}} \gamma_5 \psi_e)~,
\label{hamil_eq}
\end{eqnarray}
which gives a decay rate 
\begin{eqnarray}
\frac{d^3 \omega}{dp d\Omega_e d\Omega_{\nu}} \propto
F(Z,p)p^2 E_{\nu}^2 (1+a\frac{p}{E}\cos \theta_{e \nu}+b\frac{m}{E}) \nonumber 
\\  \times \frac{M_f}{M_i-E + p \cos \theta_{e \nu}}~,
\label{eq: distribution}
\end{eqnarray}
where $E$, $p$ and $m$ are the total energy, momentum and mass of the 
positron, $E_\nu$ the energy of the neutrino, $M_i$ and $M_f$ are the masses
of the parent atom and daughter nucleus, and 
$F(Z,p)$ the Fermi function.
We assume that the Standard Model provides an exact 
description of the
$W^{\pm}$ exchange process and that $C_V=C_V^{\prime}$\cite{note2}, but
make no assumptions on the parity or time-reversal properties of
the scalar interaction. Then the $e$-$\nu$ correlation coefficient,
\begin{equation} 
a= \frac{2-|\tilde{C}_S|^2-|\tilde{C}_S^{\prime}|^2+2 Z \alpha m/p \:
\text{Im} (\tilde{C}_S+\tilde{C}_S^{\prime})}
{2+|\tilde{C}_S|^2+|\tilde{C}_S^{\prime}|^2}~,
\label{eq: a}
\end{equation}
and the Fierz interference coefficient,
\begin{equation} 
b= -2\sqrt{1-(Z\alpha)^2} ~ \frac{\text{Re}[\tilde{C}_S+\tilde{C}_S^{\prime}]}
{2+|\tilde{C}_S|^2+|\tilde{C}_S^{\prime}|^2}~,
\label{eq: b}
\end{equation} 
are functions of $\tilde{C}_S$ and $\tilde{C}_S^{\prime}$ where
\begin{equation}
\tilde{C}_S = \frac{C_S}{C_V}~~~~\text{and}
 ~~~~~\tilde{C}_S^{\prime} = \frac{C_S^{\prime}}{C_V}~.
\end{equation}
The $\tilde{C}$'s will be complex if the scalar sector violates
time-reversal invariance.

The $e$-$\nu$ correlation must be inferred from the recoil
momentum of the daughter nucleus. This
traditionally was measured directly, restricting 
high-precision results to a few favorable cases such as 
$^6$He\cite{jo:63} and neutron decay\cite{st:78}. 
However, if the daughter nucleus is particle unstable, 
the daughter momentum can be determined from its decay products. 
This allows one to study energetic light particles rather
than slow heavy ions whose atomic and even chemical effects must
be considered. Furthermore the transformation from the rest frame of the
daughter to the lab amplifies the lepton recoil effects by a factor
$2 V/v$ where $V$ is the center-of-mass velocity of the light particle
and $v$ is the daughter's veleocity due to lepton recoil.
Finally, the time scale for particle decay is so short
that the delayed particle is emitted before the recoiling daughter 
can slow down appreciably.
Clifford et al.\cite{cl:89} compared the energies of delayed $\alpha$'s
detected in coincidence with $\beta$'s emitted toward and away 
from the $\alpha$, while Schardt and Riisager\cite{sc:93} studied the 
broadening of narrow delayed proton groups due to lepton recoil. 
The coincidence technique allows
one to measure a {\em shift} rather than a {\em spread} which is favorable
on statistical grounds and is less sensitive to the response of the
charged-particle detector. On the other hand, the extracted value of 
the correlation coefficient is very sensitive to the energy and angle 
of the detected $\beta$ particle.
We adopted the singles technique because it seemed difficult to determine
the beta's kinematics with sufficient precision.
\begin{figure}
\vspace{6cm}
\caption{Intrinsic shapes of the $0^+ \rightarrow 0^+$ 
delayed proton group for
$a=+1,~b=0$ (heavy curve) and $a=-1,~b=0$ (light
curve). The daughter's 20 eV natural width is not
visible on this scale.}
\label{fig: recoil line shape}
\end{figure}
\vspace{-11.8cm}
\hbox{\hspace{-4cm}
\psfig{figure=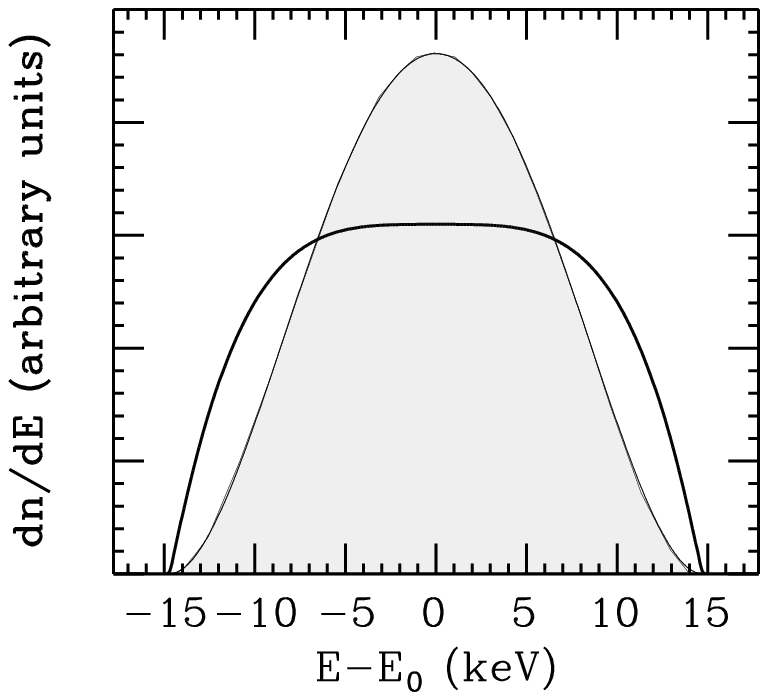,height=22cm}}
\vspace{-10.5cm}

The kinematics of $^{32}$Ar superallowed decay determine the daughter's
velocity distribution and thereby the broadening
of the delayed proton peak. The maximum kinetic energy of the
recoiling $^{32}$Cl nucleus is 
\begin{equation}
T_{\text{max}}= \frac{\Delta^2-m^2}{2M_i}
\end{equation}
where $\Delta$ is the difference in atomic masses of the parent and
daughter states.
The accepted value of $\Delta$
is poorly known because of the  $\pm 50$~keV uncertainty in the $^{32}$Ar 
mass\cite{au:95}. We use the isospin-multiplet mass equation\cite{an:85},
$M(T_3)=c_0 +c_1 T_3 +c_2 T_3^2$,
to obtain an improved value for $\Delta$ from the measured masses
of all 5 members of the $A=32$, $T=2$ multiplet.
These masses, shown in Table~\ref{tab: IMME}, were obtained from the
known ground-state masses and excitation energies\cite{en:90}
of the isobars, except for $^{32}$Cl which we computed
from our measured energy of the superallowed delayed-proton
peak in $^{32}$Ar decay, $E_{LAB}=3349.9 \pm 1.2$ keV~\cite{note1} 
and the known proton and $^{31}$S masses. 
The isospin-multiplet mass equation provides an excellent
fit to the data and predicts that
$\Delta = 3c_2-c_1=6087.3 \pm 2.2$~keV\cite{note3}, implying 
$T_{\text{max}} \approx 638$~eV
and a maximum daughter velocity of $v=2.07 \times 10^{-4} c$. 
The $^{32}$Cl daughter
state has a width $\Gamma \approx 20$~eV (see below) so 
that in one mean life
the daughter travels at most 
$2.1\times 10^{-2}$\AA~before emitting the proton.
The recoiling $^{32}$Cl therefore emitted the proton while it was still
traveling with the full velocity it received from lepton
recoil.
The intrinsic shape of the delayed proton peak
(the shape for a counter with perfect energy resolution)
is shown in Fig.~\ref{fig: recoil line shape} for the limiting cases 
$a=+1,~b=0$ and $a=-1,~b=0$. 

We performed our experiment at ISOLDE. Beams of 60~keV
$^{32}$Ar and $^{33}$Ar ions from the General Purpose
Separator were focused 
through a 4 mm diameter collimator and implanted in
a 22.7~$\mu$g/cm$^2$ carbon foil inclined at $45^{\circ}$ to the 
beam axis. Protons were detected in a pair of 9 mm $\times$ 9 mm 
PIN diode detectors collimated by 7.72 mm $\times$
7.72 mm apertures located 1.6~cm from the beam axis.
We eliminated possible uncertainties
from beta summing effects by placing the detection apparatus inside a
3.5~T superconducting solenoid. The magnetic field
prevented the betas from reaching the 
proton detectors (the highest energy betas from the 
$0^+ \rightarrow 0^+$ decay had $R_c=0.53$~cm), but had little effect 
on the protons (the superallowed proton group 
had $R_c=7.56$~cm).

The PIN diodes were maintained at $-11$ C
by thermoelectric elements that held the diode temperatures constant to 
$\pm 0.02$ C. The signals were amplified by 
preamplifiers located immediately outside the vacuum
chamber. The preamplifier housings were held at $+20$ C by 
thermoelectric devices that held the housing temperatures constant to 
$\pm 0.01$ C. Condensation of vacuum system
contaminants on the detectors and stopper foil was minimized by surrounding 
them with a copper shield cooled by a steady flow of liquid nitrogen.
As an added precaution, the detectors were warmed to 
$+27^{\circ}$ C once each day to drive off any condensed material. 
The preamplifier signals were amplified and digitized by modules mounted in
temperature-controlled crates and recorded in event-mode by a 
mini-computer. For each event we recorded the detector
energy signals, the absolute time, the delay time
after the arrival of a proton pulse, and the temperatures of 
the detectors, preamps, NIM crate, liquid nitrogen shroud,
and the room. Our system gave excellent resolution; the pulser peaks for the
two detectors had full-widths at half-maximum of 2.98 and 3.27~keV.

Data were taken over a period of 12 days under several different conditions:
with the stopper foil at $45^{\circ}$, 
$135^{\circ}$, $225^{\circ}$ and $315^{\circ}$ with respect to the
beam axis, and for two different beam tunes. These produced 6 different
spectra for each of the 2 counters. We continually
alternated between $\approx2$ h long $^{32}$Ar runs and 5-15 min long 
$^{33}$Ar runs that provided energy calibrations for the $^{32}$Ar
data. The $^{32}$Ar and $^{33}$Ar beam intensities on 
target, 
\begin{table}
\caption{Comparison of the measured mass excesses of the lowest $T=2$
quintet in $A=32$ to predictions of the Isobaric Multiplet Mass
Equation [$P(\chi^2, \nu)=0.71$].}
\label{tab: IMME}
\begin{tabular}{lrrr}
isobar & $T_3$ & $M_{\text{exp}}$~(keV)\tablenote{unless noted otherwise,
ground state masses are from Ref.~\protect\cite{au:95}}  & $M_{\text{IMME}}$
(keV) \\
\tableline
$^{32}$Si        & $+2$ & $-24080.9 \pm 2.2$ & $-24081.9 \pm 1.4$ \\
$^{32}$P         & $+1$ & $-19232.88 \pm 0.20$\tablenote{$E_x=5072.44 \pm 0.06$
keV from Ref.~\protect\cite{en:90}} & $-19232.9 \pm 0.2$ \\
$^{32}$S         & $0$ & $-13970.98 \pm 0.41$
                             \tablenote{$E_x=12045.0 \pm 0.4$~keV
from Refs.~\protect\cite{an:85,wa:98}} & $-13971.1 \pm 0.4$ \\
$^{32}$Cl        & $-1$ & $-8296.9\pm 1.2$\tablenote{from delayed
proton energy\protect\cite{note1} and masses of
Ref.~\protect\cite{au:95}.} & $-8296.6 \pm 1.1$ \\
$^{32}$Ar        & $-2$ & $-2180 \pm 50$ & $-2209.3 \pm 3.2 $ \\
\end{tabular}
\end{table}
averaged over the entire run, were 94 and 3900 ions/s.

\begin{figure}
\vspace{7cm}
\caption{Fit (upper panel) and residuals (lower panel)
of the $0^+ \rightarrow 0^+$ delayed proton peak. This spectrum,
(the sum of detector 2 data in reflection geometry) contains roughly 1/4 of
our data. The energy scale is 0.500 keV/channel. 
The pulser peak 
shows the electronic resolution.
The Breit-Wigner tail 
from the 20 eV daughter width is visible on the high-energy side 
of the peak.}
\label{fig: 1998 fits}
\end{figure}
\vspace{-11.2cm}
\hbox{\hspace{-1cm}
\psfig{figure=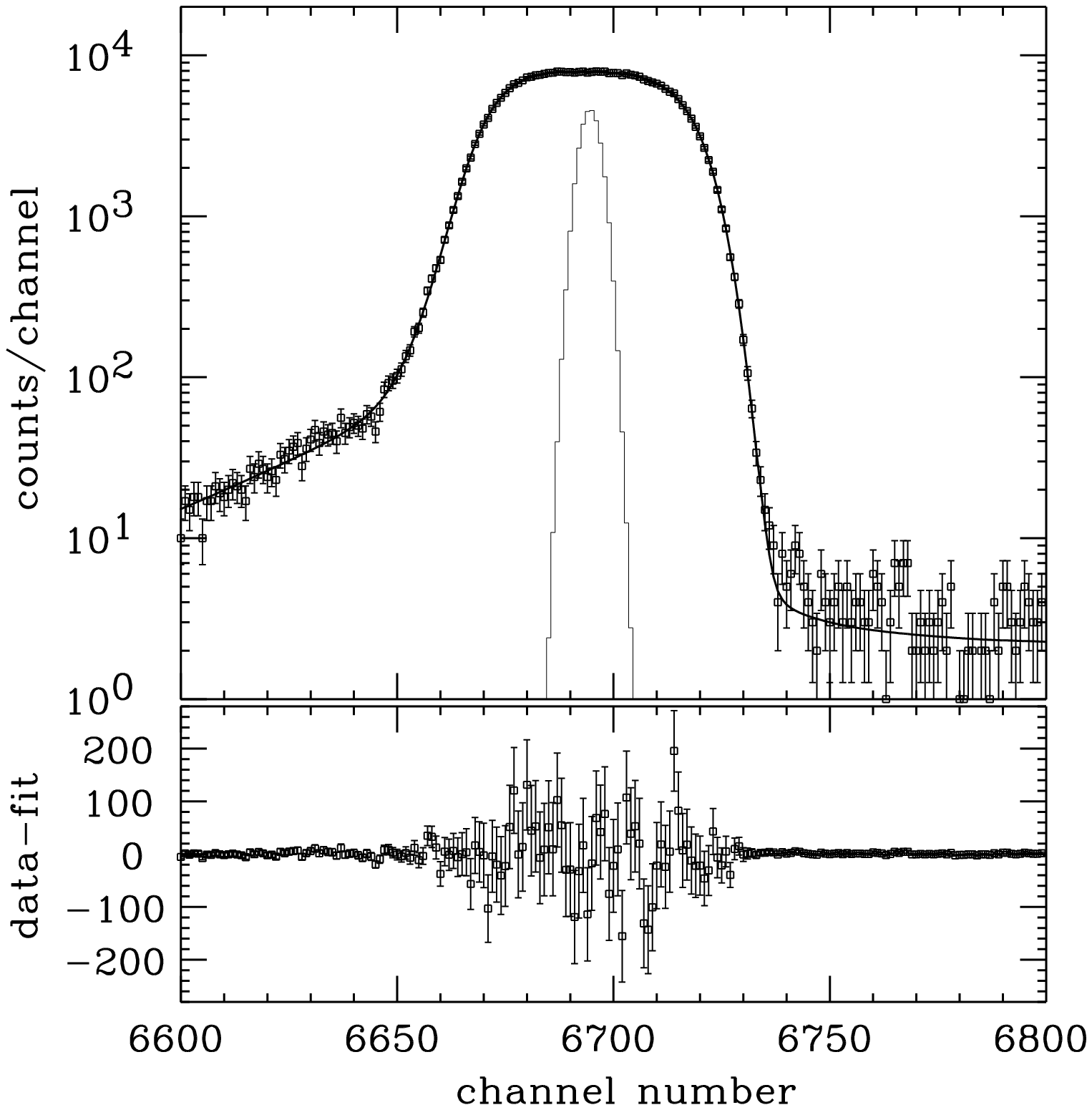,height=11cm,width=11cm}}
%
We computed the intrinsic proton shapes using Monte Carlo 
routines to generate $\beta$-decay events distributed according to
Eq.~\ref{eq: distribution} with
\begin{equation} 
E_{\nu} = \frac{(E_{\text {m}}-E)}{1+(p \cos \theta_{e\nu}-E)/M_i}~~~~~~
E_{\text {m}} = \Delta-T_{\text{max}}~.
\end{equation}
The Fermi function for a screened, finite-sized nuclear 
charge was interpolated from Tables II and III of Ref.~\cite{be:69},
and Gl\"uck's\cite{gl:98} order-$\alpha$ radiative 
correction to the energy distribution of recoiling $^{32}$Cl nuclei
was applied.
A predicted\cite{go:71} $6.7 \times 10^{-4}$ electron-capture branch 
was also included.
Protons were ejected isotropically in the $^{32}$Cl frame and deflected by the
magnetic field; the mean energy losses of individual protons
in the stopper foil and detector dead layer
(roughly 1.5 and 1.8 keV, respectively) were computed.
The stopper foil thickness was deduced from the energy loss of
3183~keV $^{148}$Gd $\alpha$'s in the foil,
while the implantation profile of Ar ions in the foil was computed
with TRIM\cite{TRIM}. The detector dead layer thicknesses, 
$23.4 \pm 0.4$ and $21.6 \pm 0.7~\mu$g/cm$^2$, were measured by
inserting into the apparatus a jig that allowed the 
$^{148}$Gd source to be moved along an arc centered on one of 
the PIN detectors.
The dead-layer loss varied as the secant of the angle while the energy 
loss in the source was essentially constant because the $\alpha$ source 
was always perpendicular to the line of sight to the detector.

We evaluated the intrinsic proton shape at 961 points on 
a $\tilde{C}_S$-$\tilde{C}_S^{\prime}$ grid. The intrinsic shapes plus a small,
flat background were convoluted with 
a proton detector response function
consisting of 2 low-energy exponential tails folded with a 
Gaussian. This functional form gave a good parameterization of
a ``first principles'' calculation of the response function that included
the effects of pulser resolution, the Fano factor, 
electronic and nuclear straggling in the stopper foil and 
detector dead layer, escape of Si X-rays, and energy 
loss to phonon excitations of the detector.
We fitted our 6 pairs of delayed-proton spectra by varying 
the response function parameters (Gaussian width, 
tail lengths and fractional areas)
to minimize $\chi^2$ for each $\tilde{C}_S$, $\tilde{C}_S^{\prime}$ point. 
Figure~\ref{fig: 1998 fits} displays the quality of the fits.
Figure~\ref{fig: lineshape comparison} shows that the extracted 
lineshape agreed well with the
``first-principles'' calculation. Fits were made for a series of
values of $\Gamma$, the natural width of the daughter.
There was essentially no correlation between $\Gamma$ and 
$\tilde{C}_S$ and $\tilde{C}_S^{\prime}$,
and we found $\Gamma = 20 \pm 10$~eV.
\begin{figure}
\vspace{7cm}
\caption{Comparison of the detector response function extracted from
the data in Fig.~\ref{fig: 1998 fits}
to the ``first-principles'' calculation described in the text.}
\label{fig: lineshape comparison}
\end{figure}
\vspace{-11.5cm}
\hbox{\hspace{-3cm}
\psfig{figure=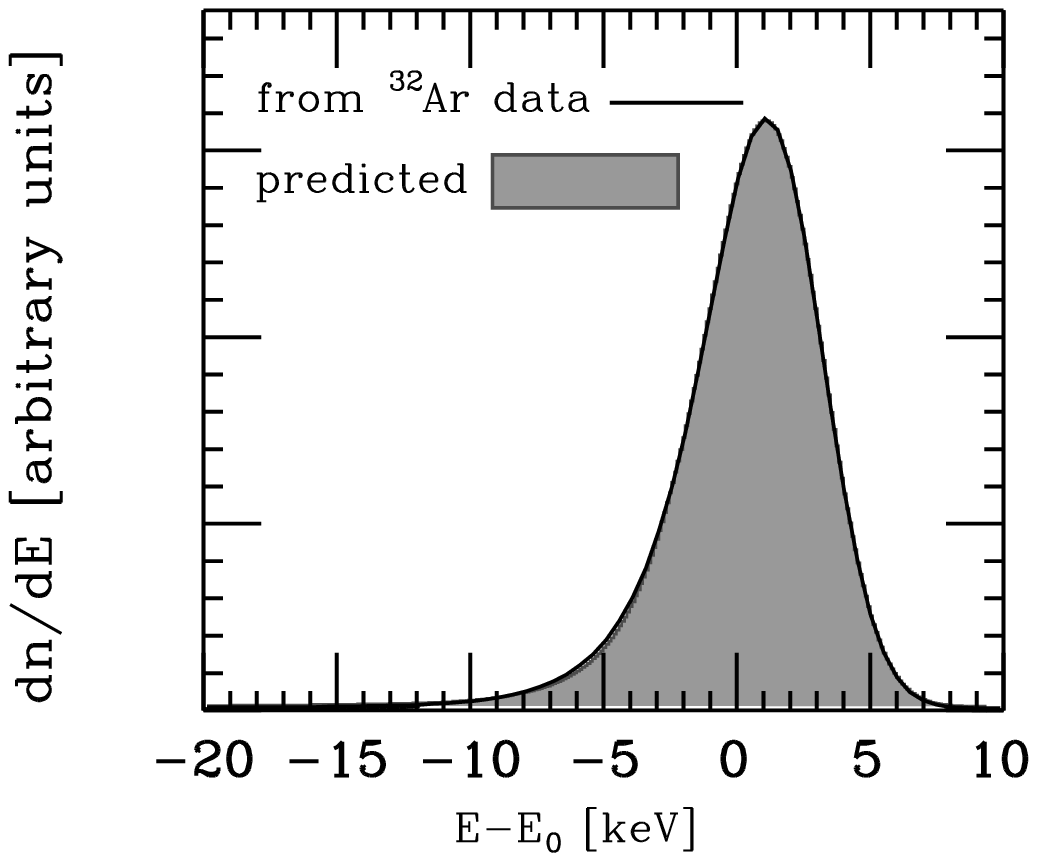,height=20cm,width=16cm}}
\vspace{-8cm}

Figure~\ref{fig: constraints} shows constraints on $\tilde{C}_S$ and 
$\tilde{C}_S^{\prime}$ from this work and from 
Refs.~\cite{or:89,pdg:98,ca:91,we:68,sc:83}. 
The annular shapes of our constraints arise from the Fierz
interference term in Eq.~\ref{eq: distribution}.
Our $\tilde{C}_S$-$\tilde{C}_S^{\prime}$ constraints may be parameterized as
\begin{eqnarray}
\tilde{a} &\equiv& a/(1 + 0.1913 b)~ \nonumber \\
&=&0.9989 \pm 0.0052({\rm stat.}) \pm 0.0036({\rm syst.})
~{\rm 68\%~c.l.}
\label{eq: result}
\end{eqnarray}
where $a$ and $b$ are given in Eqs.~\ref{eq: a},
\ref{eq: b} with 
$\langle m/p \rangle = 0.21$. Note that $\tilde{a}$, unlike $a$, does not
have an upper bound of $+1$, so the range spanned by
our experimental $2\sigma$ error band lies entirely within the physical region.

The systematic error included in Eq.~\ref{eq: result} and 
Fig.~\ref{fig: constraints} was evaluated by combining in quadrature the 
following effects.
We found the dependence of $\tilde{a}$ on the exact values of 
$\Delta$ and $Q_p$, 
$\partial \tilde{a}/\partial \Delta = -1.2\times 10^{-3}$~keV$^{-1}$ and 
$\partial \tilde{a}/\partial Q_p=-0.9 \times 10^{-3}$~keV$^{-1}$,
by repeating the entire analysis
with $\Delta$ and $Q_p$ changed by $\pm 10$~keV. 
The uncertainties, 
$\delta \Delta=\pm 2.2$~keV and $\delta Q_p= \pm 1.2$~keV, gave a
{\em kinematic} systematic error $\delta \tilde{a} = \pm 0.0032$.\cite{note4} 
We checked the dependence of $\tilde{a}$ on the {\em fitting regions} of
the proton spectra; a 28\% variation in the 
width of the region changed $\tilde{a}$ by less than $\pm 0.00055$.
We examined the dependence of our results
on the form of the detector response function by reanalysing
the data with a single-tail response function; 
by reanalyzing the data assuming that
a weak Gamow-Teller peak lay under the tail of the $^{32}$Ar superallowed
peak; 
and by simultaneously fitting the
$^{33}$Ar and $^{32}$Ar superallowed peaks using a common response
function. From these tests we inferred a
{\em lineshape} systematic error of $\delta\tilde{a} = \pm 0.0016$.

\vspace{-1.5cm}
\hbox{\hspace{-3cm}
\psfig{figure=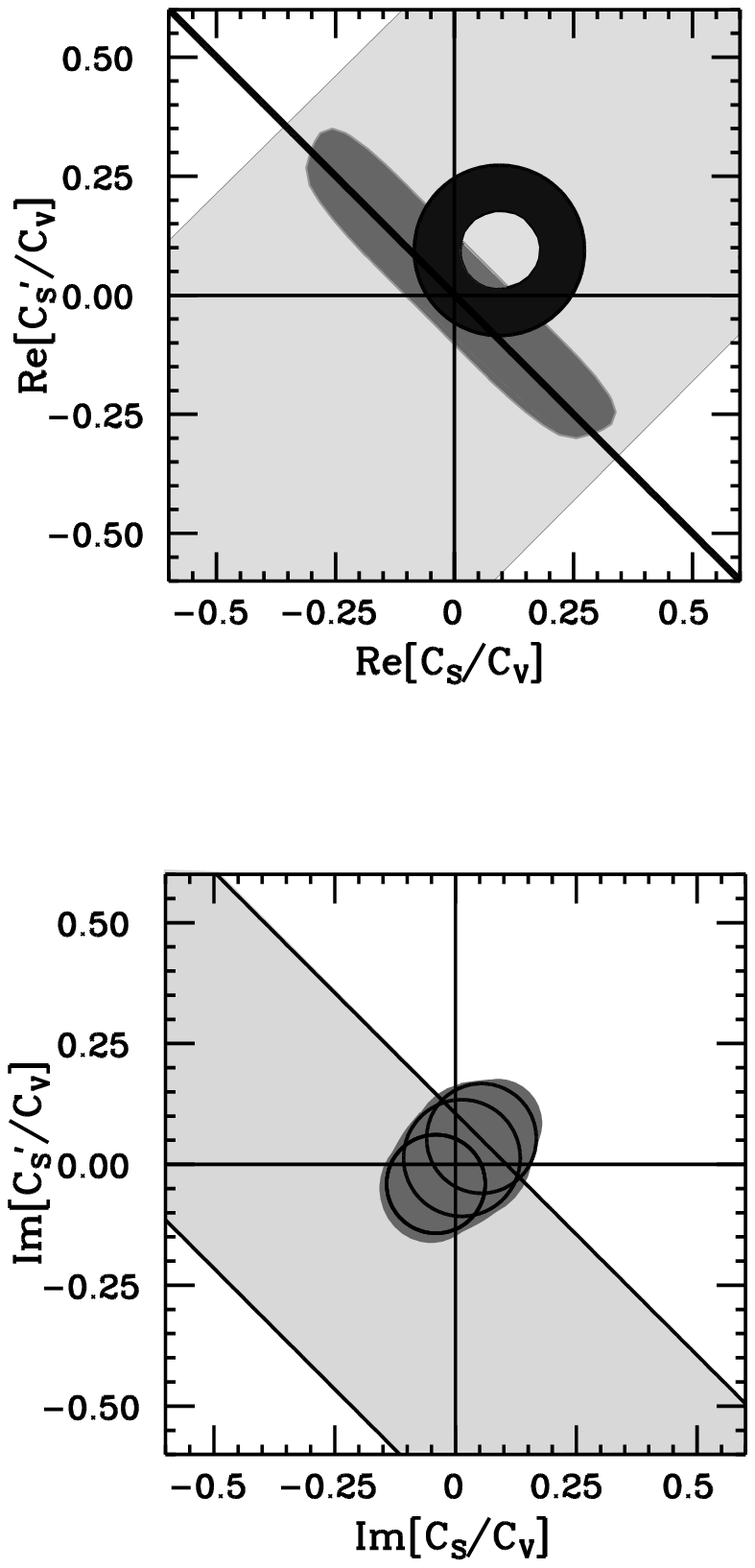,height=20cm,width=16cm}}
\vspace{-7.0cm}
\begin{figure}
\vspace{1.0cm}
\caption{95\% conf. limits on $\tilde{C}_S$ and $\tilde{C}_S^{\prime}$.
Upper panel: time-reversal-even couplings. 
The annulus is from this work. The narrow diagonal band is from
$b(0^+ \rightarrow 0^+)$~\protect\cite{or:89}. 
The broad diagonal band shows constraints from 
$A$, $B$, $a$, and $t_{1/2}$ in $n$ decay\protect\cite{pdg:98}; 
the sausage-shaped area includes, in addition,
constraints from $G(^{14}$O) and $G(^{10}$C) \protect\cite{ca:91},
$b(^{22}$Na) \protect\cite{we:68} and $a(^6$He) \protect\cite{jo:63}.
Lower panel: time-reversal-odd
couplings. The circles are from this work and correspond to 
$\tilde{C}_S$ and $\tilde{C}_S^{\prime}$ phases of
$\pm 90^{\circ}$, $+45^{\circ}$ and $-45^{\circ}$. The 
shaded oval shows the constraint with no assumptions about this phase.
The diagonal band is from $R(^{19}$Ne) \protect \cite{sc:83}.}
\label{fig: constraints}
\end{figure}
For scalar interactions with 
$\tilde{C}_S=-\tilde{C}_S^{\prime}$ so that $b=0$,
we obtain a $1\sigma$ limit
$|\tilde{C}_S|^2 \leq 3.6 \times 10^{-3}$. The corresponding
limit on the mass of scalar particles with gauge coupling strength
is  
$M_S = |\tilde{C}_S|^{-1/2} M_W \geq 4.1 M_W$. 

We thank T. Van Wechel and H. Simons for help constructing the apparatus,
J.J. Kolata and F. Bechetti for lending us the superconducting
magnet, D. Forkel-Wirth for help setting up our apparatus at CERN,
and B. Jennings and I. Bigi for useful remarks. 
This work was supported in part by the DOE 
(at the University of Washington) 
and by the NSF and the Warren Foundation 
(at the University of Notre Dame). 

\end{narrowtext}

\begin{references}
\bibitem{note2}We neglect possible small leptoquark-exchange contributions
to $C_V$ and $C_V^{\prime}$.
\bibitem{note1}The error is dominated by the uncertainties in
the energies of the $^{33}$Cl daughter states fed in $^{33}$Ar decay.
\bibitem{note3}There is no evidence for any violation of the IMME, 
$P(\chi^2,\nu)$ drops to 0.52 when a $c_3T_3^3$ term is added.
\bibitem{note4}We account for the correlation between $\Delta$ and 
$Q_p$.
\end{references}
\end{document}